\documentclass[twocolumn]{jpsj2}
\usepackage{amsmath}
\usepackage[dvips]{graphicx,color}
\makeatletter

\@addtoreset{equation}{section}
\makeatother

\def\runauthor{Online-Journal Subcommittee}
\title{
Pressure dependence of the resistivity  around  valence transition
\\ based on $1/N$-expansion study for extended  Anderson lattice
}
\author{%
Yasutaka \textsc{Nishida}\thanks{E-mail address: 
y-nishida@blade.mp.es.osaka-u.ac.jp} and Tomokazu Ookubo
\thanks{E-mail address: 
ookubo@blade.mp.es.osaka-u.ac.jp} }
\inst{%
Department of Materials Engineering Science,\\
Graduate School of Engineering Science, Osaka University, \\
Toyonaka, Osaka 560-8531
}
\recdate{\today}
\abst{%
We investigate the behavior of the resistivity around valence transition
in the one-dimensional extended periodic Anderson model (PAM) with the
Coulomb repulsion between f and conduction electron $U_{\textrm{fc}}$.
By using $1/N$-expansion in the leading order in $(1/N)^0$, 
where $N$ is the spin-orbital degeneracy
of f-electrons, valence transition happens at critical $U_{\textrm{fc}}$
with increasing the atomic f-level as seen in the previous work \cite{Oni}.
We  calculate the resistivity whole temperature range around valence transition
based on the extended PAM including the crystal field
and investigate how the physical properties such as Kondo temperature is modified by 
additional $U_{\textrm{fc}}$ interaction.
As a result, the double peak structure of the electrical resistivity fades away rapidly by 
the rapid increase of the Kondo temperature.
}
\kword{%
electrical resistivity, valence transition, 
1/$N$-expansion, Extended Anderson lattice}
\begin{document}
\maketitle
%%%%%%%%%%%%%%%%%%%%%%%%%%%%%%%%%%%%%%%%%%%%%%%%
%               INTRODUCTION                   %
%%%%%%%%%%%%%%%%%%%%%%%%%%%%%%%%%%%%%%%%%%%%%%%%
\section{Introduction}
Heavy Fermion Systems have been investigated extensively both experimentally and theoretically
as typical system of strongly correlated electrons.
In 1979 Steglich and collaborators discovered an intermetallic of the rare earth metal cerium
$\textrm{Ce}$$\textrm{Cu}_2$$\textrm{Si}_2$\cite{steg}, which is well known as first discovered
superconductor with $T_\textrm{c}\sim 0.7\textrm{K}$ at ambient pressure.
In the phase diagram in $P$-$T$ plane of $\textrm{Ce}$$\textrm{Cu}_2$$\textrm{Si}_2$ under the high pressure,
two superconducting transition temperature is confirmed.
The isostructual compound, $\textrm{Ce}$$\textrm{Cu}_2$$\textrm{Ge}_2$\cite{jac,jac2}, also has the same
phase diagram.

These two compounds have been considered to have the  similar physical properties
if the origin of the pressure for $\textrm{Ce}$$\textrm{Cu}_2$$\textrm{Si}_2$
is shifted.
According to the recent experimental research\cite{Holmes}, the mechanism of the the superconductivity at higher 
pressure region is seemed to be different from the ordinary antiferromagnetic spin fluctuation,
because transition temperature is away from the point of magnetic instability at $T=0$.
Around the superconducting transition temperature at higher pressure $P_c$, 
the residual resistivity has a peak, and the coefficient of $T^2$-term of the resistivity shows 
rapid decrease.
At high temperature, 
two temperatures corresponding to maxima of the magnetic resistivity coincides around $P_c$.
From these results, these coincidence around $P_c$  seemed to be  related to a  rapid change of valence of Ce\cite{Oni,Mae}, 
so that the superconductivity mediated by the valence fluctuation was proposed.

In this short paper, we study the pressure dependence of the resistivity 
based on the pioneering work\cite{Oni} on the valence fluctuation by the extended
periodic Anderson model with f-c coulomb repulsion $U_{\textrm{fc}}$.
Then, we use $1/N$-expansion\cite{Ono1,Ono2,Nishi} for extended Anderson lattice including 
crystal field, calculate both the low temperature region  ($T\le T_K/10$, where 
$T_K$ is the Kondo temperature.)
and the high temperature region ($T\ge T_K$) within $(1/N)^0$.
In the case of $U_{\textrm{fc}}=0$, the double peak structure due to the crystal field is shown to merge
into single peak gradually with increasing $\varepsilon_f$$(<0)$.
On the other hand, in the case of $U_{\textrm{fc}}\ne0$,
we find that the double peak structure fades away more rapidly due to the increase of $T_K$ caused
by rapid change of $n_f$.  
%Experimentally around $P_c$, the rapid decrease of the Kadowaki-Woods ratio is also observed,
%the rapid change of $T_K$ caused by $U_{\textrm{fc}}$ term is 
%
%

The paper is organized as follows. In $\S 2$, we introduce the model Hamiltonian and $1/N$-expansion.
In $\S$ 3 and 4, we show the result the physical properties both low and high temperature region, and 
the pressure dependence of the resistivity. 
In section 5, we summarize conclusions and remarks.
\section{Model and Formal Preliminaries}
In this paper, we start with an extended periodic Anderson model, which is a periodic 
Anderson model (PAM) with f-electron in a manifold of $J=5/2$\cite{yama1}, and Coulomb repulsion
between f and conduction electrons, $U_{\textrm{fc}}$: Our model Hamiltonian is given by
\begin{align}
H&=H_{\textrm{c}} + H_{\textrm{f}} + H_{\textrm{hy}} + H_{\textrm{fc}},\\
H_{\textrm{c}}&=\sum_{{\bf k}\sigma}\varepsilon_{{\bf k}\sigma}c^{\dag}_{{\bf k}\sigma}c_{{\bf k}\sigma},\\
H_{\textrm{f}}&=\sum_{i\Gamma}E_{i\Gamma}f_{i\Gamma}^{\dag}f_{i\Gamma},\\
H_{\textrm{hy}}&=\frac{1}{\sqrt{N_L}}\sum_{\sigma,i,{\bf k},\Gamma}(V_{{\bf k}\Gamma \sigma}\textrm{e}^
{-\textrm{i}{\bf k}\cdot {\bf R}_i}c_{{\bf k}\sigma}^{\dag}f_{i\Gamma}b^{\dag}_{i}+\textrm{h.c.}),\\
H_{\textrm{fc}}&=U_{\textrm{fc}}\sum_{i,\Gamma}n^{\textrm{f}}_{i\Gamma} n^{\textrm{c}}_i,
\end{align}
where, $c_{{\bf k}\sigma}^{\dag}$ is the creation operator for the conduction electron with wave vector ${\bf k}$
and spin $\sigma$, and $\varepsilon_{{\bf k}\sigma}$ is the energy of the conduction electron with wave vector ${\bf k}$
and spin $\sigma$. $N_L$ denotes the number of the lattice site, and $i$ stands for the site index.
Furthermore we introduce the slave-boson which represents the $f^0$-state, and pseudo-fermion which represents 
$f^1$-state, following the Coleman: $b^{\dag}_{i}$ is the creation operator for the slave-boson at $i$-site, $f^{\dag}_{i\Gamma}$
is the creation operator for the pseudo-fermion at $i$-site with Crystalline-Electric-Field (CEF) level $|\Gamma\rangle$.
The mixing potential $V_{{\bf k}\Gamma \sigma}$ is expressed as $V_{{\bf k}\Gamma \sigma}=\sum_{M}O_{\Gamma M}V_{{\bf k}M\sigma}$,
$M$ is the z-component of angular momentum $J$ of an f-electron: $M=J_z (J=5/2)$. Using the spherical harmonic function $Y_{l}^{m}$,
$V_{{\bf k}M\sigma}$ is given by 
\begin{align}
V_{{\bf k}M\sigma}&=V_0{\sqrt{\frac{4\pi}{3}}} \Bigl\{ -2\sigma \sqrt{\frac{(\frac{7}{2}-2M\sigma)}{7}} Y_{l=3}^{M-\sigma}(\theta_{ k},\phi_{{ k}})  \Bigr\}.
\end{align}
$O_{\Gamma M}$ is determined by the symmetry of the crystal field.
In tetragonal symmetry we can put each f-level as follows,
\begin{align}
E_{i|\pm 1\rangle}=\varepsilon_{f}+\Delta_2,\ 
E_{i|\pm 2\rangle}=\varepsilon_{f}+\Delta_1,\  
E_{i|\pm 3\rangle}=\varepsilon_{f}.
\end{align}
where, the CEF splitting is parameterized by $\Delta_{1,2}$.

\begin{figure}
\begin{center}
\includegraphics[scale=0.35]{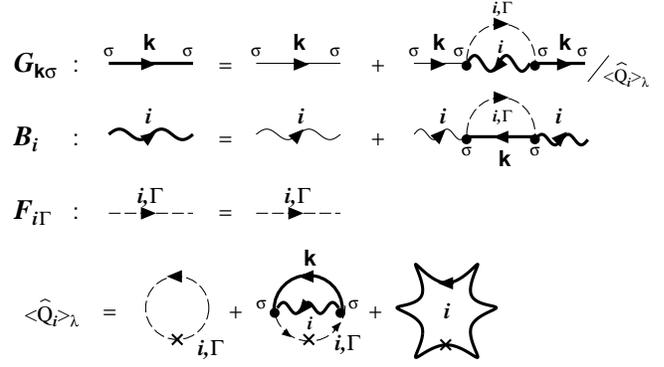}
\end{center}
\caption{Diagrammatic representation of Dyson equations for single Green function within
accuracy of $(1/N)^0$.  Solid line denotes the conduction electron propagator, 
wavy line  the slave-boson propagator and dashed line  the pseudo-electron propagator, respectively. }
\label{fig1}
\end{figure}

Here the on-site Coulomb repulsion $U$ between f-electrons is assumed to be infinite, so that
we must impose the following local constraints in order to guarantee an equivalence between the present model Hamiltonian,
\begin{eqnarray}
\hat{Q}_i=b_{i}^{\dag}b_{i}+\sum_{\Gamma}f_{i\Gamma}^{\dag}f_{i\Gamma}=1.
\label{local}
\end{eqnarray}
The expectation value of an operator $\hat{O}$ under the local constraint eq. (\ref{local}) is given as\cite{Col}
\begin{eqnarray}
\langle \hat{O} \rangle = \lim_{ \{ \lambda_i \} \rightarrow \infty  }
\langle \hat{O} \prod_i\hat{Q}_i \rangle_{\lambda} /\langle  \prod_i\hat{Q}_i \rangle_{\lambda},
\label{expe}
\end{eqnarray}
where $\langle \hat{A}  \rangle_{\lambda}$ is calculated in the grand canonical ensemble
for the Hamiltonian $H_{\lambda}$: 
%
%-------------- (2.11)-(2.12) ---------------
%
\begin{eqnarray}
&&\langle \hat{A}  \rangle_{\lambda} \equiv \textrm{Tr}\Bigl[\textrm{e}^{-\beta H_{\lambda}}\hat{A}\Bigr]
/\textrm{Tr}\Bigl[\textrm{e}^{-\beta H_{\lambda}}\Bigr],\\
&& H_{\lambda}=H+\sum_{i}\lambda_i \hat{Q}_i.
\label{mf}
\end{eqnarray}
The single-particle Green function $G_{{\bf k}\sigma}$ for the conduction electron, $B_i$ for the slave
boson and $F_{i\Gamma}$ for the pseudo-fermion is given as follows:
\begin{align}
\label{gk}
G_{{\bf k}\sigma}(\textrm{i}\omega_n)&=(\textrm{i}\omega_n-\varepsilon_{{\bf k \sigma}}
-\Sigma_{{\bf k}\sigma}(\textrm{i}\omega_n))^{-1},\\
B_{i}(\textrm{i}\nu_n)&=(\textrm{i}\nu_n-\lambda_i-\Pi_i(\textrm{i}\nu_n))^{-1},\\
F_{i \Gamma}(\textrm{i}\omega_n)&=(\textrm{i}\omega_n -E_{i \Gamma} -\lambda_i)^{-1}.
\end{align}
%
%
%These Green's function are illustrated in Fig. \ref{fig1}.
The self-energies $\Sigma_{{\bf k}\sigma}$ and  $\Pi_i$ is calculated by $1/N$-expansion procedure.
We perform an extended calculation on the basis of the modified version of 
$1/N$-expansion method\cite{Nishi}.
In the lowest order, the Dyson equations for one-particle Green's function are
given by the diagram shown in Fig. \ref{fig1}.
Then the analytic form of the self-energy is given by
\begin{align}
\Sigma_{{\bf k}\sigma}(\textrm{i}\omega_n)=&
\lim_{\lambda_i \rightarrow \infty}\Bigl[-\sum_{\Gamma}|V_{{\bf k}\Gamma \sigma}|^{2} T\sum_{\nu_n}
B_i(\textrm{i}\nu_n)\nonumber\\
&\times F^{0}_{i \Gamma}(\textrm{i}\omega_n+\textrm{i}\nu_n)/\langle \hat{Q}_i \rangle_{\lambda}\Bigr],\\
\Pi_{i}(\textrm{i}\nu_n)=&\frac{1}{N_L}\sum_{\Gamma} \sum_{\sigma}\sum_{{\bf k}_\sigma}|V_{{\bf k}\Gamma \sigma}|^{2}
T\sum_{\omega_n} G_{{\bf k}\sigma}(\textrm{i}\omega_n)\nonumber\\
&\times F^{0}_{i\Gamma}(\textrm{i}\omega_n+\textrm{i}\nu_n),\\
\langle \hat{Q}_{i}\rangle_{\lambda}=&\sum_{\Gamma}\langle \hat{n}_{fi\Gamma}\rangle_{\lambda}+\langle \hat{n}_{bi}\rangle_{\lambda}.
\label{self}
\end{align}
where
\begin{align}
\langle \hat{n}_{fi\Gamma}\rangle_{\lambda}=\langle f_{i\Gamma}^{\dag}f_{i\Gamma}\rangle_{\lambda},
\end{align}
and
\begin{align}
\langle \hat{n}_{bi}\rangle_{\lambda}=\langle b_{i}^{\dag}b_{i}\rangle_{\lambda}.
\end{align}
By solving the coupled self-consistent equations eqs. (\ref{gk})-(2.14) over the
whole temperature range, we can discuss the temperature dependence of the physical quantities.
In the next section, we discuss the physical properties at $T=0$ as first step.
%
%-------------- Fig.2 ---------------
%
\section{Physical properties at low temperature limit}
At lower temperatures than the coherence temperature $T_0$, the coupled self-consistent equations 
become more simplified, with the help of the following relation,
\begin{align}
-\frac{1}{\pi}\textrm{Im}B_i(\omega+\textrm{i}0_+)\simeq a \delta(\omega-\lambda_i-\varepsilon_f+E_0)+C(\omega),
\label{b0}
\end{align}
where, $E_0$ and $a$ are the binding energy and residue of the slave-boson, respectively,
and $C(\omega)$ is continuum part, which is non-zero $\omega \ge \lambda_i + \varepsilon_f$.
$E_0$ corresponds to the Kondo temperature $T_K$,
and the coherence temperature $T_0$ is estimated with $E_0/10$.
$E_0$ and $a$ are determined by the relation as
\begin{align}
E_0=&\varepsilon_f - \textrm{Re}\Pi_i(\lambda_i+\varepsilon_f-E_0)\\
\frac{1}{a}=&1-\frac{\textrm{d}}{\textrm{d}\omega}(\textrm{Re}\Pi_i(\omega))\Bigr|_{\omega=\lambda_i+\varepsilon_f-E_0}
\label{e0}
\end{align}

Here we treat $U_\textrm{fc}$ term with the Hartree-Fock approximation in $O((1/N)^0)$\cite{hira}.
By using eq. (3.1), the system is in the coherent regime with the renormalized bands given by
\begin{align}
G_{{\bf k}\sigma}(\textrm{i}\omega_n)=\sum_{j=1}^{4}\frac{\tilde{A}_{{\bf k}\sigma}^j}{\textrm{i}\omega_n-\tilde{\alpha}^j_{{\bf k}\sigma}}
\end{align}
with
\begin{align}
\tilde{\varepsilon}_{{\bf k}\sigma}&=\varepsilon_{{\bf k}\sigma}+U_{\textrm{fc}}n_f,\\
{\tilde{\alpha}_{{\bf k}\sigma}^{j}}&=\tilde{\varepsilon}_{{\bf k}\sigma}+\textrm{Re}\Sigma_{{\bf k}\sigma}
({\tilde{\alpha}_{{\bf k}\sigma}^{j}}),\\
\tilde{A}_{{\bf k}\sigma}^{j}&=\frac{(\tilde{\alpha}_{{\bf k}\sigma}^j-E_0)(\tilde{\alpha}_{{\bf k}\sigma}^j-E_0-\Delta_1)(\tilde{\alpha}_{{\bf k}\sigma}^j-E_0-\Delta_2)}
{(\tilde{\alpha}_{{\bf k}\sigma}^j-\tilde{\alpha}_{{\bf k}\sigma}^{j+1})(\tilde{\alpha}_{{\bf k}\sigma}^j-\tilde{\alpha}_{{\bf k}\sigma}^{j+2})
(\tilde{\alpha}_{{\bf k}\sigma}^j-\tilde{\alpha}_{{\bf k}\sigma}^{j+3})},
\end{align}
where, ${\tilde{\alpha}_{{\bf k}\sigma}^{j}}={\tilde{\alpha}_{{\bf k}\sigma}^{j+4}}$.

Hence the self-consistent equations eqs. (2.12)-(2.14) are written by
\begin{align}
\label{selfe}
&E_0-\varepsilon_f= \frac{1}{N_L}\sum_{j=1}^{4}\sum_{\Gamma}\sum_{\sigma}\sum_{{\bf k}_\sigma}
\frac{\tilde{A}_{{\bf k}\sigma}^j |V_{{\bf k} \Gamma \sigma}|^2 f(\tilde{\alpha}_{{\bf k}\sigma}^{j})}{E_0+\tilde{E}_{i\Gamma}-\varepsilon_f-\tilde{\alpha}_{{\bf k}\sigma}^{j}},\\
&\frac{1}{a}=1+\frac{1}{N_L} \sum_{j=1}^{4}\sum_{\Gamma}\sum_{\sigma}\sum_{{\bf k}_\sigma}
\frac{\tilde{A}_{{\bf k}\sigma}^j |V_{{\bf k} \Gamma \sigma}|^2 f(\tilde{\alpha}_{{\bf k}\sigma}^{j})}{(E_0 +\tilde{E}_{i\Gamma}-\varepsilon_f - \tilde{\alpha}_{{\bf k}\sigma}^{j})^2},\\
&n=n_c+n_f=\frac{1}{N_L}\sum_{j=1}^{4}\sum_{\sigma}\sum_{{\bf k}_\sigma} f(\tilde{\alpha}_{{\bf k}\sigma}^{j})\tilde{A}_{{\bf k}\sigma}^{j} + (1-a).
\end{align}
and
\begin{align}
\tilde{E}_{i\Gamma}&=E_{i\Gamma}+U_{\textrm{fc}}n_c,
\end{align}
where $n$, $n_c$, and $n_f$ is the total electron number, the number of the conduction electrons, and f-electrons per site, respectively.
Finally, the binding energy $E_0$, the residue $a$ and other physical properties at $T=0$ are obtained from a series of equations eqs. (\ref{selfe})-(3.10).
\subsection{f-electron number}
\begin{figure}[h]
\begin{center}
\includegraphics[scale=0.4]{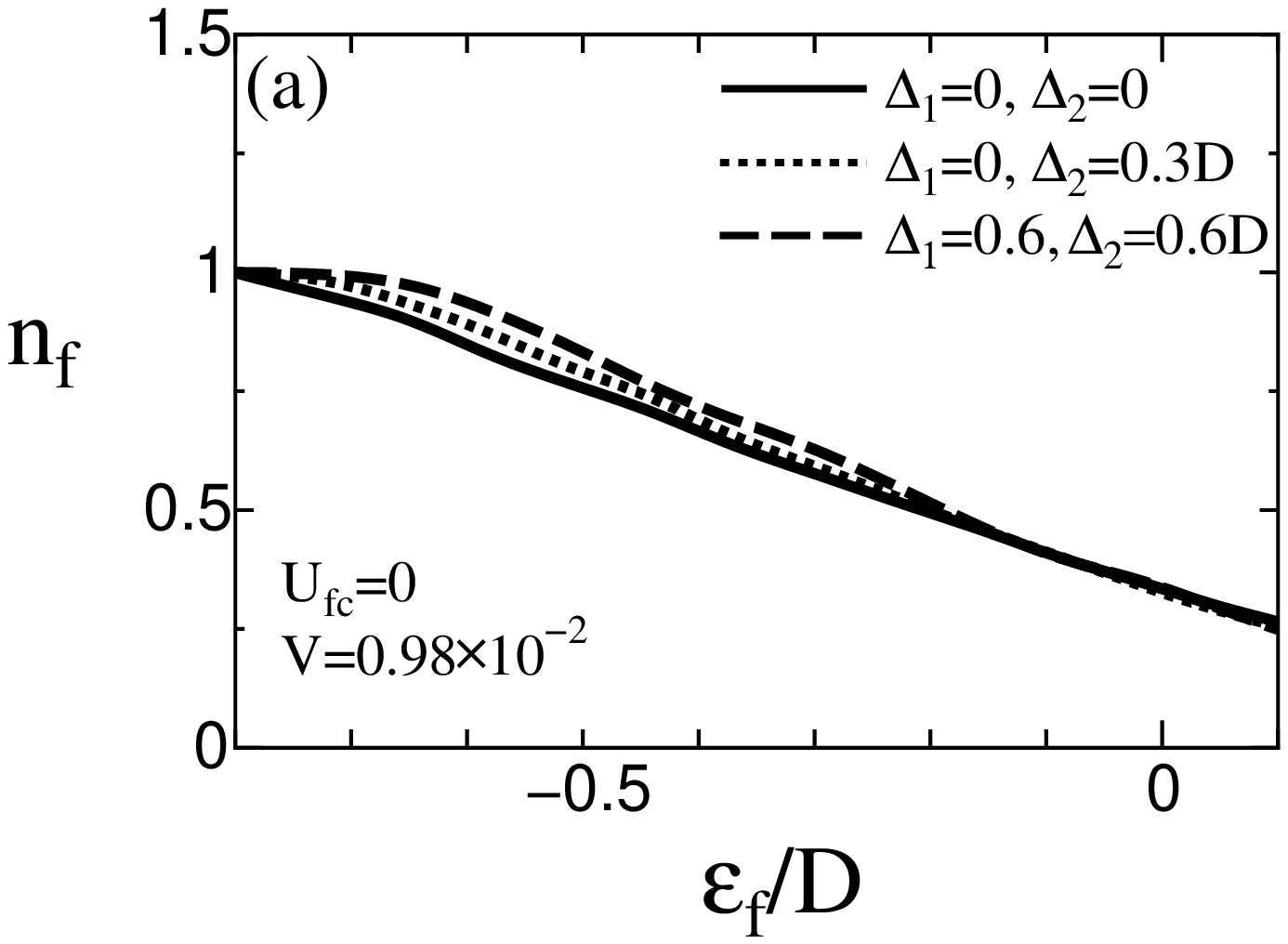}
\includegraphics[scale=0.4]{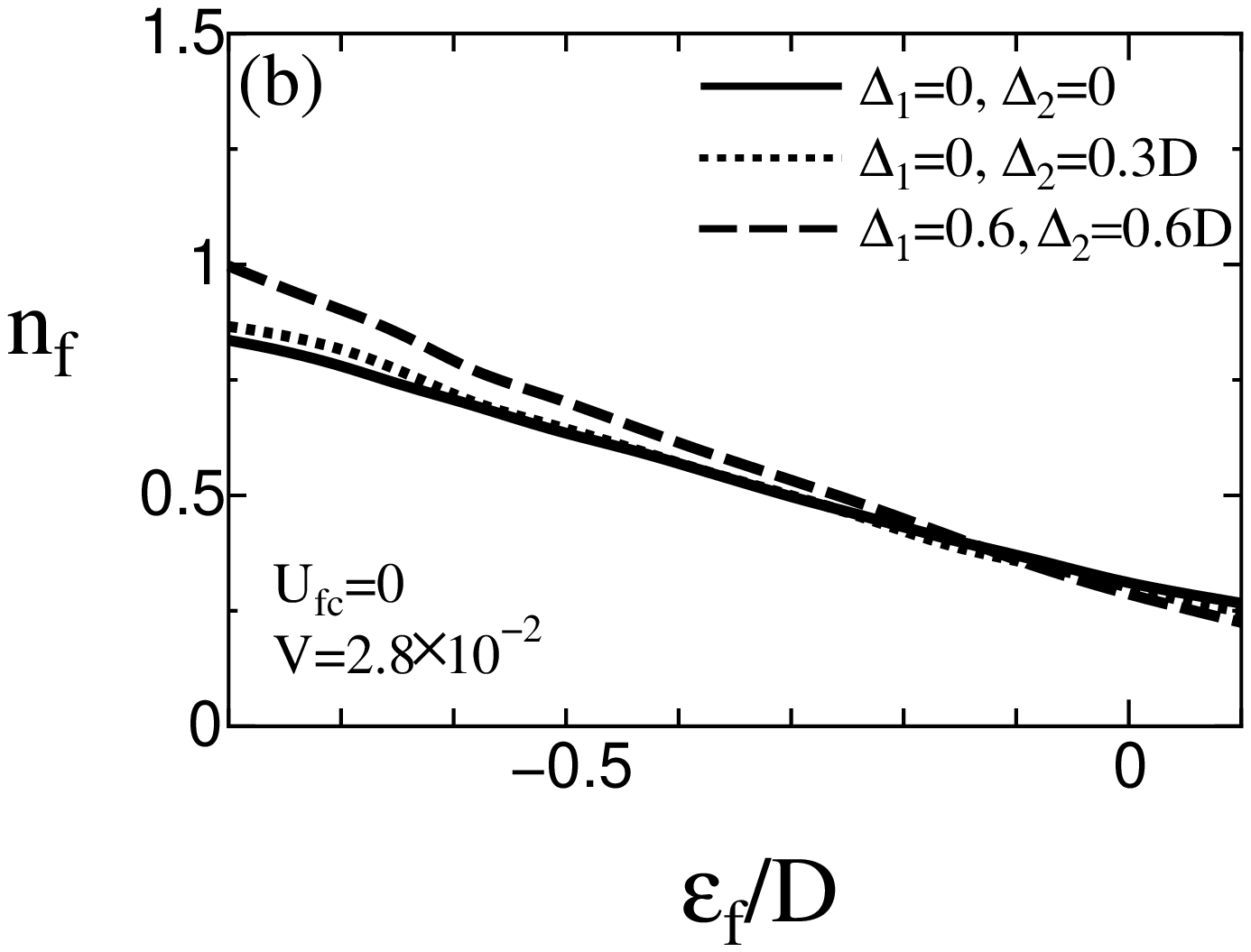}
\end{center}
\caption{(a) $n_f$ vs $\varepsilon_f$ in the case of $n=1.4$, $U_{\textrm{fc}}=0$ and $V^2=0.98\times 10^{-2}D^2$
for a series of CEF splittings.
(b) $n_f$ vs $\varepsilon_f$ in the case of $n=1.4$, $U_{\textrm{fc}}=0$ and $V^2=2.8\times 10^{-2}D^2$
for a series of CEF splittings.}
\label{nf1}
\end{figure}

In this section, we show the f-electron number $n_f$ per site  as a function of the atomic f-level 
$\varepsilon_f$ for series of CEF splitting scheme with $U_{\textrm{fc}}$.
At first, let us investigate the trivial case, 
$V^2 \equiv \sum_{\sigma}|V_{{\bf k}\Gamma\sigma}|^2 =0.98\times 10^{-2}D^2$, $n=n_c+n_f=1.4$, and $\Delta_1=\Delta_2=0.6D$.
Then  the grand state is doublet ($N_{\textrm{GS}}=2$) state, so that, in Fig. \ref{nf1}, we find that if $\varepsilon_f \lesssim -0.6$,
$n_f \sim 1$. However in case of $V^2=2.8\times10^{-2}$, $n_f$ decreases  monotonously by increasing 
 the width of the density of state of localized f-electrons due to hybridization effect.
On the other hand, in the case of $\Delta_1=\Delta_2=0$ ($N_{\textrm{GS}}=6$), 
$n_f$ decreases  monotonously. 
%By considering the CEF splittings,
%we find that the $\varepsilon_f$-dependence of  $n_f$ changes with the degeberacy
%of the grand state $N_{\textrm{GS}}$. 
If $N_{\textrm{GS}}$ is small, $n_f$ shows a little sudden decrease with increasing $\varepsilon_f$. 

\begin{figure}[h]
\begin{center}
\includegraphics[scale=0.45]{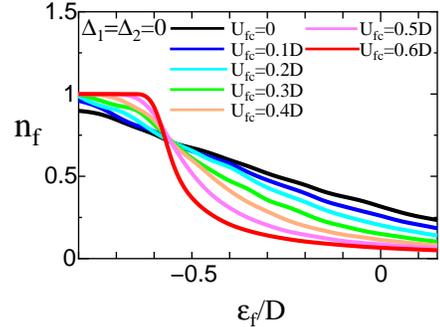}
\end{center}
\caption{$n_f$ vs $\varepsilon_f$ in the case of $n=1.4$, $V^2=2.0\times 10^{-2}D^2$
and $\Delta_1=\Delta_2=0$
for a series of $U_{\textrm{fc}}$. }
\label{nf2}
\end{figure}

\begin{figure}[h]
\begin{center}
\includegraphics[scale=0.4]{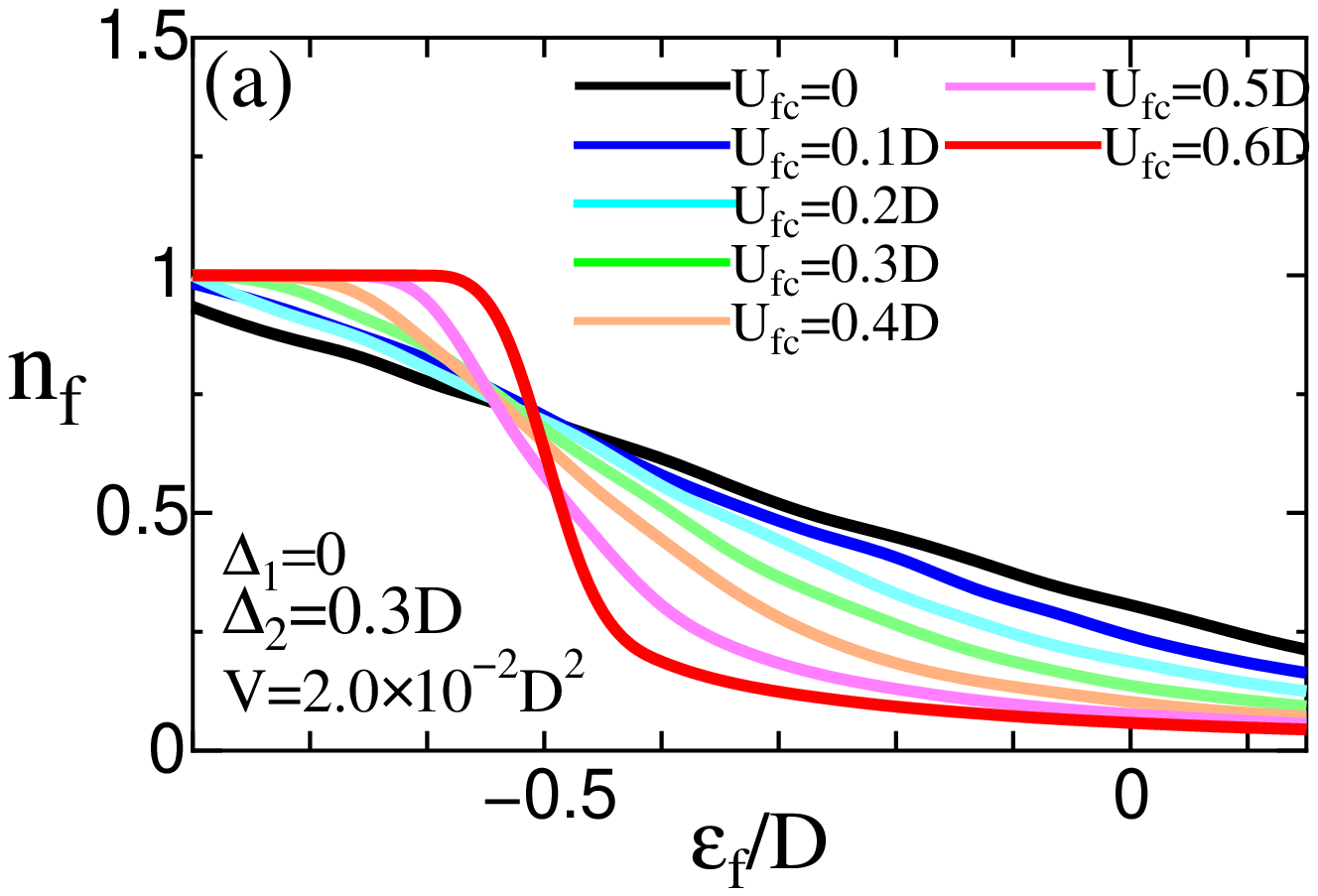}
\includegraphics[scale=0.4]{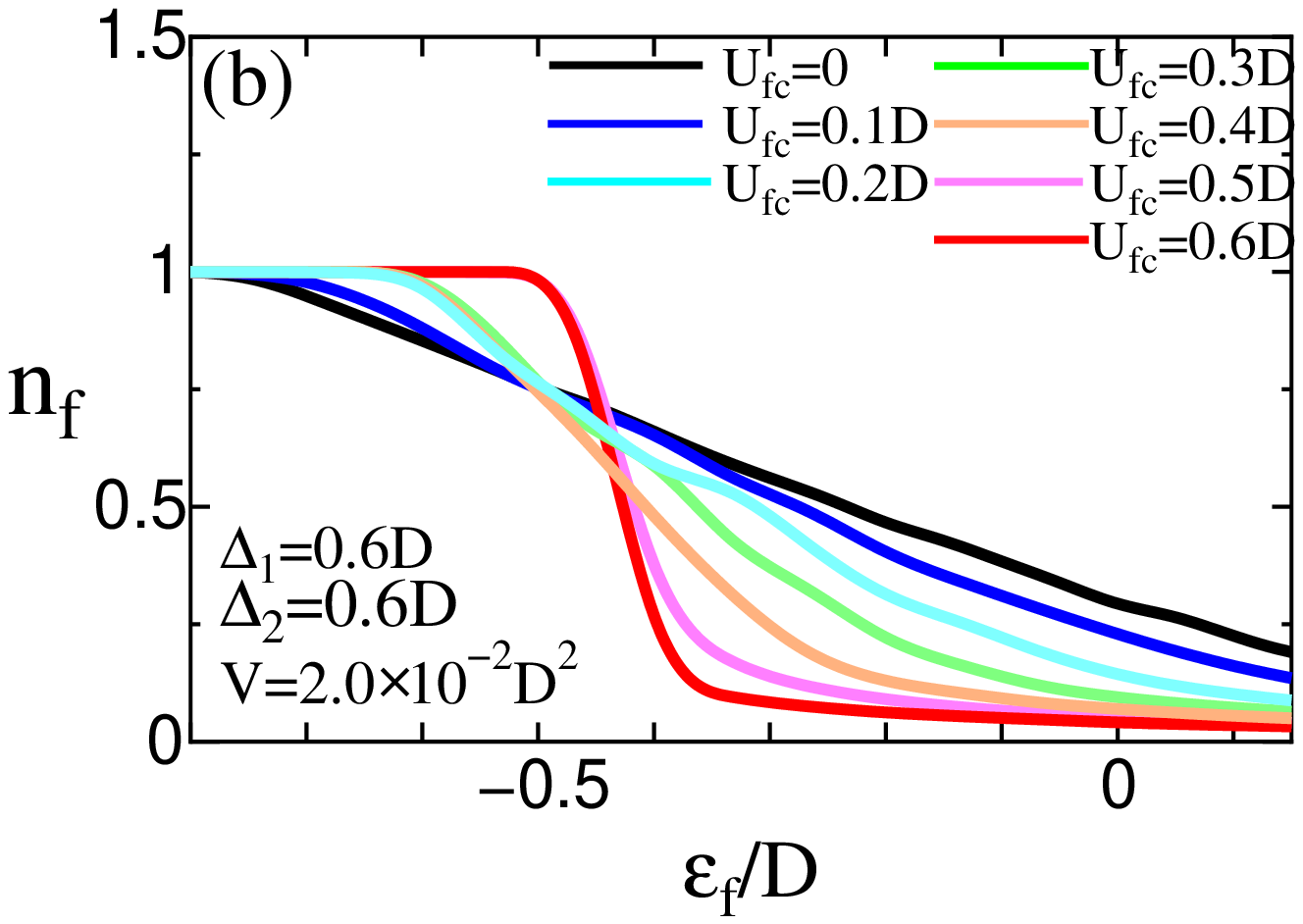}
\end{center}
\caption{(a) $n_f$ vs $\varepsilon_f$ in the case of $n=1.4$, $V^2=2.0\times 10^{-2}D^2$,
$\Delta_1=0$, and $\Delta_2=0.3D$
for a series of $U_{\textrm{fc}}$. 
(b) $n_f$ vs $\varepsilon_f$ in the case of $n=1.4$, $V^2=2.0\times 10^{-2}D^2$,
$\Delta_1=0.6D$, and $\Delta_2=0.6D$
for a series of $U_{\textrm{fc}}$. }
\label{nf3}
\end{figure}

Next, we show the $U_{\textrm{fc}}\ne 0$ result  in Fig. {\ref{nf2}} and {\ref{nf3}}.
In Fig. {\ref{nf2}}, we  set $\Delta_1=\Delta_2=0$  for simplicity.
Then, $n_f$ shows the rapid decrease gradually with increasing $U_{\textrm{fc}}$.
This behavior was already obtained by the past pioneering work\cite{Oni}.
In Fig. {\ref{nf3}}, we  set $\Delta_1=0, \Delta_2=0.3D$  and $\Delta_1=\Delta_2=0.6D$
in order to consider the CEF effect. From these result we find $n_f$ vs $\varepsilon_f$ is not much
modified qualitatively by CEF effect. 
%
%
%-------------- Fig.3---------------
%
%
%
%
%-------------- Fig.4 ---------------
%
\subsection{Kondo temperature}
The Kondo temperature corresponds to the binding energy of the slave-boson $E_0$ in lattice case.
We present the solution $E_0$ for the series of the self-consistent equations (3.8)-(3.10)
within the $O((1/N)^0)$.  
Incidentally $E_0$ can be written as follows in the case of impurity without CEF effect and $U_{\textrm{fc}}$,
\begin{align}
E_0=D \exp\Bigl(\frac{\varepsilon_f}{6\rho_0 V^2}\Bigr),
\label{kont}
\end{align}
where we impose the condition that $D \gg |\varepsilon_f| \gg E_0$.
In eq. (\ref{kont}), 6 means the degeneracy of f-electron in a manifold $J=5/2$, and $\rho_0$ represents
for the density of state of the conduction electron per spin.

\begin{figure}[h]
\begin{center}
\includegraphics[scale=0.58]{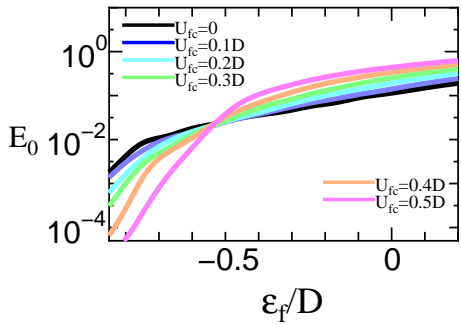}
\end{center}
\caption{$E_0$ vs $\varepsilon_f$ in the case of $n=1.4$, $V^2=2.0\times 10^{-2}D^2$,
$\Delta_1=\Delta_2=0$
for a series of $U_{\textrm{fc}}$. }
\label{konzu1}
\end{figure}

In Fig. \ref{konzu1}, we show the result of $E_0$ vs $\varepsilon_f$ for the various $U_\textrm{fc}$
and CEF splitting. At first we again consider the trivial case $\Delta_1=\Delta_2=0$ and $U_{\textrm{fc}}=0$.
Then, $E_0$ increase gradually with increasing $\varepsilon_f(<0)$. This tendency is expected from the impurity
result eq. (\ref{kont}).
Next, in case of $U_{\textrm{fc}}\ne 0$, $E_0$ changes drastically as large as $n_f$ with increasing $\varepsilon_f$.
This is explained by the enhancement of the renormalization factor $q$ accompanying with the rapid valence change. 
The renormalization factor $q$ derived from mean-field\cite{gut} and Variational Monte Calro\cite{gut2} is obtained as follows
\begin{align}
q^{-1}=\frac{1-n_f/2}{1-n_f}.
\end{align} 
The rapid change $n_f: 1\rightarrow 0$ causes also the rapid enhancement of $q$, so that
the Kondo temperature is enhanced around valence transition.

\begin{figure}[h]
\begin{center}
\includegraphics[scale=0.55]{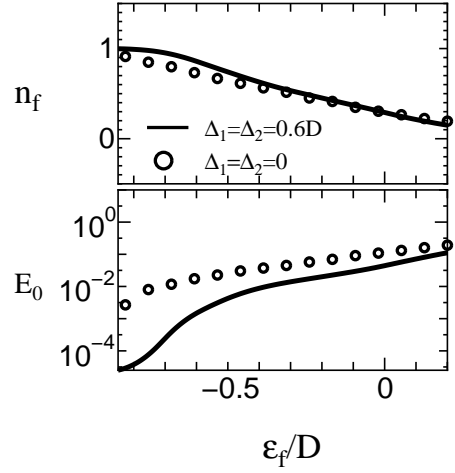}
\end{center}
\caption{$n_f$ vs $\varepsilon_f$ and $E_0$ vs $\varepsilon_f$ in the case of $n=1.4$, $V^2=2.0\times 10^{-2}D^2$.
The result of $\Delta_1=\Delta_2=0.6D$ is shown by solid line, and that of $\Delta_1=\Delta_2=0$ is shown by enclosed circle.}
\label{konzu2}
\end{figure}

In Fig. \ref{konzu2}, we illustrate the result of the $\Delta_1=\Delta_2=0.0$ and $\Delta_1=\Delta_2=0.6D$
under the condition $U_{\textrm{fc}}=0$.
In the previous subsection, we find the CEF effect does not affect the $\varepsilon_f$ dependence of $n_f$,
on the other hand, Kondo temperature $E_0$ is greatly influenced of CEF effect.  
The value of the Kondo temperature depends much by the degeneracy of the localized f-electrons.
%-------------- Fig.5 ---------------
%
\section{Electrical Resistivity}
In this section, the effect of the  pressure on the temperature dependence of the resistivity is discussed.
The pressure effect is parameterized as the atomic f-level $\varepsilon_f$ measured from the band center of conduction electrons.
The conductivity is obtained by means of the Kubo formula.
By omitting some constant factors, we define the reduced conductivity,
\begin{align}
\sigma=\frac{1}{N_L}\sum_{\sigma,{\bf k}}v_{{\bf k}}^2\int \textrm{d}\varepsilon \Bigl(
-\frac{\partial f(\varepsilon)}{\partial \varepsilon} \Bigr) \Bigl[
\textrm{Im}G_{{\bf k}\sigma}(\varepsilon+\textrm{i}0_+)
\Bigr]^2,
\end{align}
where we use the renormalized Green function derived from $1/N$-expansion procedure.
In order to calculate the resistivity $1/\sigma$ over the whole temperature range, 
we must solve the Dyson equation eqs. (2-12)-(2.14) self-consistently (self-consistent $1/N$). 

\begin{figure}[t]
\begin{center}
\includegraphics[scale=0.5]{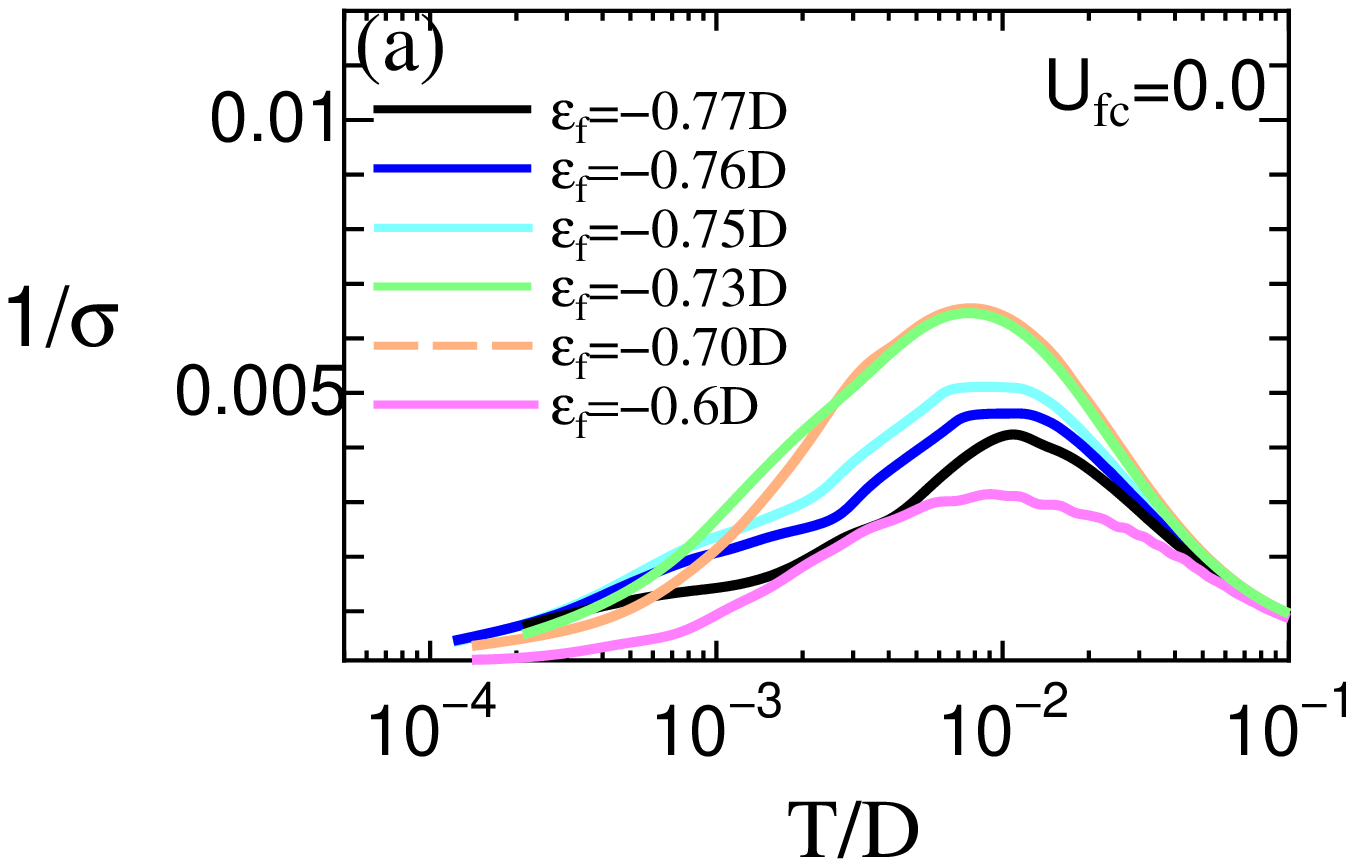}
\includegraphics[scale=0.5]{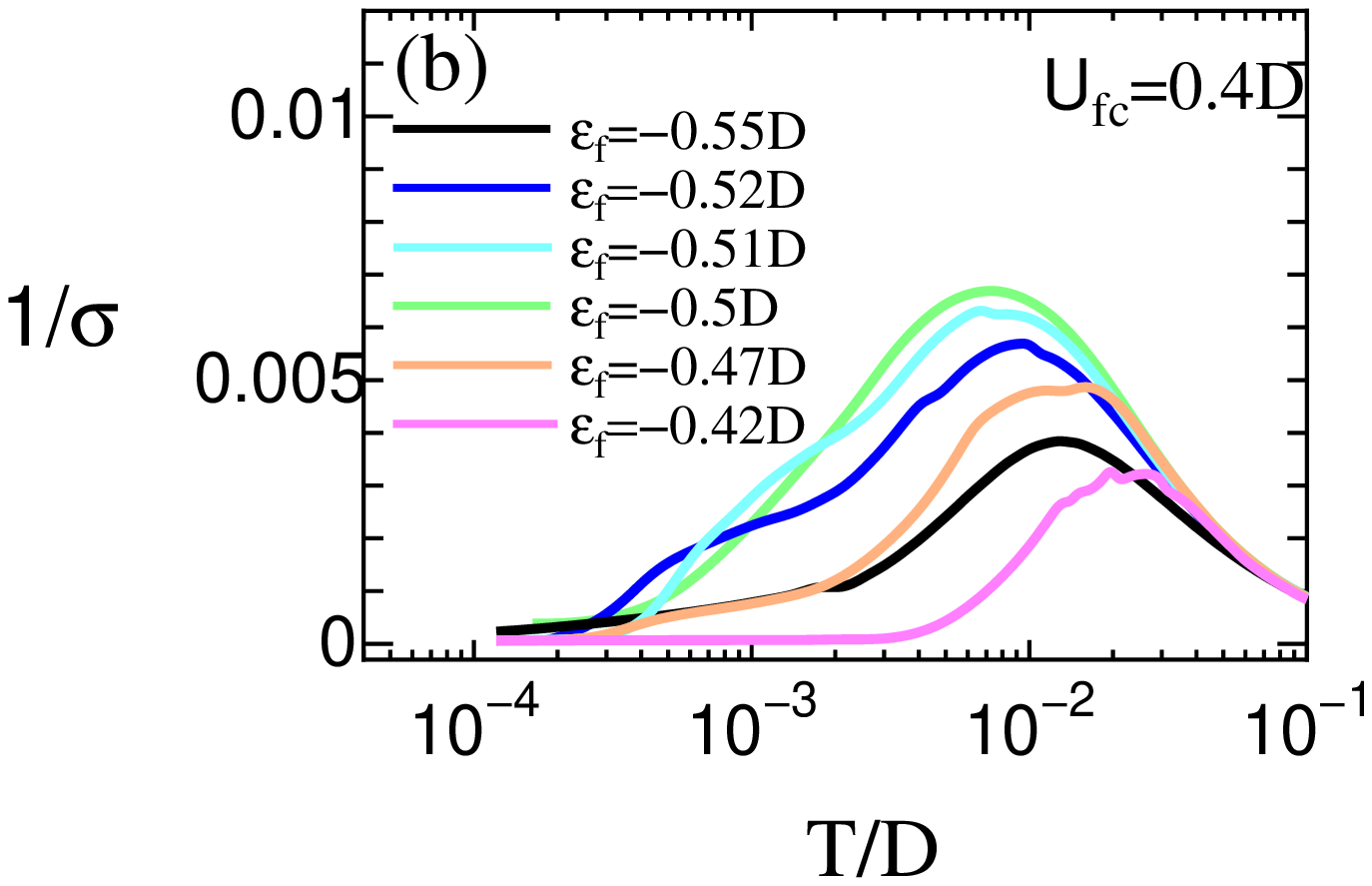}
\end{center}
\caption{(a) The temperature dependence of electrical resistivity for a series of the atomic f-level
in the case of $U_{\textrm{fc}}=0$ and (b) that in the case of $U_{\textrm{fc}}=0.4D$.
The CEF splittings are set as $\Delta_1=0.015D$ and $\Delta_2=0.038D$ in both cases. }
\label{fig6}
\end{figure}

We illustrate the result of the calculation for the series of $\varepsilon_f$ in Fig. \ref{fig6}
The parameters are adopted as $D$=1(band width), $V^2 \equiv \sum_{\sigma}|V_{{\bf k}\Gamma\sigma}|^2 =1.28\times 10^{-2}D^2$, $n=n_c+n_f=1.4$,
and set the  CEF parameter as $\Delta_1=0.015D$ and $\Delta_2=0.038D$ putting the case of $\textrm{Ce}\textrm{Cu}_2\textrm{Si}_2$\cite{Steg}.

In both case of $U_{\textrm{fc}}=0$ and $U_{\textrm{fc}}\ne 0$, the double peak structure fades away with 
increasing $\varepsilon_f(<0)$. The double peak structure of the temperature dependence is commonly shown to merge into
a single peak with increasing pressure. These behaviors are consistent with the tendency observed in Ce-based heavy fermions.
However, at the large $\varepsilon_f$ (high pressure region),
the position of the single peak of $1/\sigma$ differs much in both cases.
In case of $U_{\textrm{fc}}=0$, the peak position does not change much at large $\varepsilon_f$.
On the other hand, in case of $U_{\textrm{fc}}\ne 0$, the peak position changes much with increasing $\varepsilon_f$.
The Fig. \ref{efnf} shows this situation clearly, i.e., rapid increase of $E_0$ (Kondo temperature) happens
around the valence transition.
Under the condition without $U_{\textrm{fc}}$, the Kondo temperature increases slowly with increasing $\varepsilon_f$
compared with the case of $U_{\textrm{fc}}=0.4D$.
The effect of the additional term $U_{\textrm{fc}}$  seems to promote the merging of the double peak structure of the resistivity.

\begin{figure}[t]
\begin{center}
\includegraphics[scale=0.5]{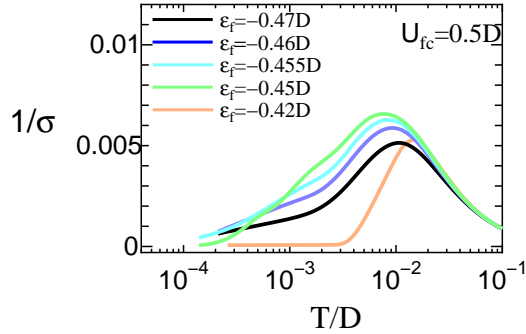}
\end{center}
\caption{The temperature dependence of electrical resistivity for a series of the atomic f-level
in the case of $U_{\textrm{fc}}=0.5D$.}
\label{ucf3}
\end{figure}

\begin{figure}[h]
\begin{center}
\includegraphics[scale=0.6]{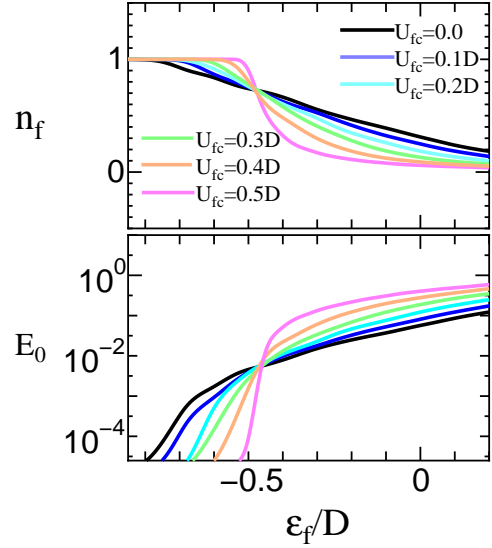}
\end{center}
\caption{$n_f$ vs $\varepsilon_f$ and $E_0$ vs $\varepsilon_f$ in the case of $n=1.4$, $V^2=1.28\times 10^{-2}D^2$,
$\Delta_1=0.015D$ and $\Delta_2=0.038D$ for the series of $U_{\textrm{fc}}$.}
\label{efnf}
\end{figure}

We also illustrate the result in case of $U_{\textrm{fc}}=0.5D$ in Fig. \ref{ucf3}.
We find that the electrical resistivity becomes more sensitive to the  change of $\varepsilon_f$ than that of $U_{\textrm{fc}}=0$. 
\section{Conclusion}
We have studied the behavior of the pressure dependence of the electrical resistivity around valence transition
based on $1/N$-expansion for extended periodic Anderson lattice including CEF effect.
This is the extended calculation of the pioneering work\cite{Oni}.
At lower temperature than Kondo temperature, the result of $n_f$ vs $\varepsilon_f$ does not change qualitatively,
i.e., $n_f$ shows the rapid decrease with increasing $U_{\textrm{fc}}$ where CEF effect does not affect the tendency 
of the change of $n_f$ as a function of $\varepsilon_f$.
remarkably.
However the Kondo temperature $T_K$ shows different behavior by CEF effect, because $T_K$ depends on the 
degeneracy of the localized f-electrons much.
Moreover, we have calculated the electrical resistivity over the whole temperature region by solving the Dyson equation
self-consistently within $(1/N)^0$ and investigated the effect of  the additional interaction $U_{\textrm{fc}}$
to the temperature dependence of the electrical resistivity.

As a result, we found that the double peak structure of the resistivity due to the crystal field fades away more rapidly
caused by the rapid change of $n_f$.
Especially, in the region of $\varepsilon_f$ around the valence transition, the resistivity show the drastic change
by a minute change of $\varepsilon_f$.
The effect of the additional term $U_{\textrm{fc}}$  promote the merging of the double peak structure of the resistivity.
This tendency may be consistent with the  merge of the resistivity right above $P_c$
in the phase diagram in $P$-$T$ plane of CeC$\textrm{u}_2$S$\textrm{i}_2$ or CeC$\textrm{u}_2$G$\textrm{e}_2$. 

Finally we point out some remaining problem and perspectives.
In our calculation, the term of $U_{\textrm{fc}}$ is treated within Hartree-Fock approximation, so that
the effect of $U_{\textrm{fc}}$ may be overestimated by the present calculation. In order to discuss the problem more
qualitatively, we need more proper approximation.
Experimentally, the rapid decrease of Kadowaki-Woods ratio is observed around $P_c$.
This suggests that the system is changed from Kondo regime to the valence fluctuation regime rapidly.
We can discuss this problem by studying up to $(1/N)^1$, where intersite correlation can be discussed.
By investigating  between the change of the degeneracy of the f-electron and pressure effect or $U_{\textrm{fc}}$,
we may also discuss the mechanisms of enhancement of $T_K$ in detail.

%
%
%%%%%%%%%%%%%%%%%%%%%%%%%%%%%%%%%%%%%%%%%%%%%%%%%%%%%%%%%%%%%%%%%%%%%%%%%%%%%%%%%%
%                               Acknowledment                                    %
%%%%%%%%%%%%%%%%%%%%%%%%%%%%%%%%%%%%%%%%%%%%%%%%%%%%%%%%%%%%%%%%%%%%%%%%%%%%%%%%%%
%
\section*{Acknowledgement}
We would like to thank A. Tsuruta and  K. Miyake for valuable discussions.
One of the authors (Y. N.)  also would like to thank K. Hattori for stimulating discussions.
This work is supported by a Grant-in-Aid for Scientific Research (No.16340103) and 21st Century
COE Program (G18) from the Japan Society for the Promotion of Science.
%
%
%
%
%%%%%%%%%%%%%%%%%%%%%%%%%%%%%%%%%%%%%%%%%%%%%%%%%%%%%%%%%%%%%%%%%%%%%%%%%%%%%%%%%%
%                                 Ref                                            %
%%%%%%%%%%%%%%%%%%%%%%%%%%%%%%%%%%%%%%%%%%%%%%%%%%%%%%%%%%%%%%%%%%%%%%%%%%%%%%%%%%
%

\end{document}